\documentstyle[aps,prl,floats]{revtex}
\begin{document}
\draft
\tighten
\twocolumn[\hsize\textwidth\columnwidth\hsize\csname@twocolumnfalse%
\endcsname

\title{TeV Neutrinos and GeV Photons from Shock Breakout in Supernovae}

\author{Eli Waxman$^\dagger$ \& Abraham Loeb$^\star$}

\address{$\dagger$ Department of Condensed Matter Physics, Weizmann
Institute, Rehovot 76100, Israel; waxman@wicc.weizmann.ac.il\\
$\star$ Astronomy Department, Harvard University, 60 Garden
Street, Cambridge, MA 02138, USA; aloeb@cfa.harvard.edu
}

\date{\today}

\maketitle

\begin{abstract}

We show that as a Type II supernova shock breaks out of its progenitor
star, it becomes collisionless and may accelerate protons to energies
$>10$~TeV. Inelastic nuclear collisions of these protons produce a $\sim
1$~hr long flash of TeV neutrinos and 10~GeV photons, about $10$~hr after
the thermal (10 MeV) neutrino burst from the cooling neutron star.
A Galactic supernova in a red supergiant star would produce a photon and
neutrino flux of $\sim 10^{-4}~{\rm erg~cm^{-2}~s^{-1}}$. A km$^2$ neutrino
detector will detect $\sim 100$ muons, thus allowing to constrain both
supernova models and neutrino properties.

\end{abstract}


]

\narrowtext

Type II supernovae are triggered by core collapse in a massive star,
generating a strong shock wave which propagates through the progenitor star
and ejects its envelope (see \cite{WoosleyWeaver86} for a review).
Interest in the breakout of the shock out of the envelope of the star, was
prompted by Colgate's suggestion that it may be accompanied by energetic
$\gamma$-ray radiation due to Bremsstrahlung and inverse-Compton emission
of thermal electrons \cite{Colgate74}. Numerical simulations
concluded that the shock
breakout should be accompanied by a burst of hard UV and X-ray radiation
\cite{Klein789,Ensman92}.

In the optically-thick interior of the progenitor, the shock is dominated
by radiation. The dissipation of kinetic energy at the shock front is
mediated by Compton scattering rather than by collisional viscosity, and
the thermal energy density behind the shock is mostly in the form of
radiation.  Numerical simulations suggest that during breakout, the
radiation-dominated shock is transformed into a viscous shock
\cite{Klein789,Ensman92}.  
When the optical depth of the gas lying
ahead of the shock is no longer very large, radiation decouples from the
gas and the shock is transformed into a collisional, viscosity--mediated
shock.

In this {\it Letter}, we show that as the shock becomes viscous,
electro-magnetic instabilities at the shock front grow at a rate which is
many orders of magnitude higher than the collision rate, leading to
scattering of ions and to magnetic field amplification on a time scale much
shorter than the viscous time scale. Under similar conditions,
astrophysical shocks are known to become collisionless (i.e., to be
mediated by collective plasma instabilities rather than by collisional
viscosity) and to accelerate charged particles to relativistic energies
(see \cite{Eichler87} for a review). We derive the expected flux and
spectrum of high energy photons and neutrinos which are produced through
inelastic nuclear collisions of the accelerated protons.

As we show below, the resolution of available numerical simulations is not
sufficient to correctly describe the evolution of plasma parameters across
the viscous shock (see also \cite{Klein789}).  We therefore first derive
the expected collisional shock structure from basic principles, and only
then discuss the stability of the shock.  While our analysis provides a
description of the shock structure, it does not allow to determine the
exact radius where the viscous shock forms.  Numerical calculations suggest
that the transition to a viscosity-dominated shock occurs at an optical
depth (measured from the stellar surface) between $\tau\sim1$
\cite{Klein789} and $\tau\sim10$ \cite{Ensman92}. We therefore parameterize
our results in terms of $\tau$. More detailed numerical simulations are
necessary to allow an accurate determination of this parameter.

We focus our discussion on stars which are believed to be the most common
progenitors of Type II supernovae, namely red-supergiants (RSG)
\cite{WoosleyWeaver86}. 
Our model predicts emission of high energy particles also
for blue-supergiants (BSGs), such as the unusual progenitor of
SN1987A. However, the predicted neutrino and $\gamma$-ray fluxes from the
surface layer of BSGs are predicted to be two orders of magnitude smaller
than from RSGs.  Additional emission may result from the interaction of the
supernova shock with the progenitor wind on long time scales \cite{wind},
although this emission depends sensitively on the wind speed and mass loss
rate and might also be strongly modulated by past variability of the wind
\cite{variability}.

\paragraph*{Physical parameters at shock breakout.}

When the radiation-dominated shock reaches an optical depth lower than
unity, radiation can no longer couple to the gas and so the shock becomes
viscous. This occurs very close to the surface radius of the progenitor
star, $R_\star$.  For the high temperatures of interest, the opacity is
dominated by Thomson scattering, and the mass of a shell of optical depth
$\tau$ located ahead of the shock is given by
\begin{equation}
M=4\pi R_*^2\frac{\mu m_p}{\sigma_T}\tau=1.8\times10^{-5}R_{*,13.5}^2
\tau~M_\odot,
\label{eq:mass}
\end{equation}
where $R_{\star,13.5}\equiv(R_\star/10^{13.5}$~cm$)=1$ reflects the
characteristic radius of a RSG progenitor. As mentioned above, we ignore
rarer, more compact progenitors, such as blue supergiants for which
$R_{\star,13.5}=0.1$, since the shell mass at a fixed
$\tau$, and hence the energy output in high-energy neutrinos and photons,
are smaller by a factor $\sim 10^2$ in them compared to RSGs.  Using the
analytic approximations of Matzner \& Mckee \cite{Matzner00} to describe
the outer envelope structure, we find that the mass density of gas at
the RSG surface is
\begin{equation}
\rho= 0.54 \times 10^{-10} \tau^{3/5}~{\rm g~cm}^{-3},
\label{eq:density}
\end{equation}
and the shell thickness is 
$\Delta=1.1\times10^{11}\tau^{2/5}~{\rm cm}$.

The shock velocity at breakout is $v_s\approx10^9{\rm cm~s^{-1}}$
(e.g. \cite{Matzner00}), corresponding to an ion temperature of
\begin{equation} 
T_i=\frac{3}{16} \mu
v_s^2=1.0\times10^5v_{s,9}^2~{\rm eV}, \label{eq:T_i} 
\end{equation} 
where $\mu=m_p/2$ is the mean atomic mass per particle for a plasma in
which the thermal energy is shared equally between protons and electrons.  If
the shock is radiation--dominated, the radiation temperature is
approximately given by the relation $aT_r^4\approx\rho v_s^2$, i.e.
\begin{equation}
T_r\approx30\rho_{-10}^{1/4}v_{s,9}^{1/2}{\rm eV}, 
\label{eq:T_r}
\end{equation} 
where $\rho_{-10}= (\rho/10^{-10}~{\rm g~cm^{-3}})$.

\paragraph*{Viscous shock structure.}

The ion-ion, ion-electron, and electron-electron collision rates are given by
\begin{equation}
\nu_{ii}\approx 3\rho_{-10}T_{i,5}^{-3/2}~{\rm s}^{-1},
\label{eq:nu_ii}
\end{equation}
\begin{equation}
\nu_{ie}\approx10^3\rho_{-10}\times
\cases{6T_{i,5}^{-3/2}~{\rm s}^{-1},&if $T_e<m_eT_i/m_p$;\cr
2T_{e,2}^{-3/2}~{\rm s}^{-1},&otherwise,\cr}
\label{eq:nu_ie}
\end{equation}
\begin{equation}
\nu_{ee}\approx 10^4\rho_{-10}T_{e,4}^{-3/2}~{\rm s}^{-1}.
\label{eq:nu_ee}
\end{equation}
Throughout this {\it Letter}, we use the notation
$T_{\alpha,x}=(T_\alpha/10^x$~eV$)$ to denote the temperature of component
$\alpha$ ($=i$ for ions, $e$ for electrons, or $r$ for radiation) 
of the plasma. 
The collisional energy loss of ions is dominated by ion-ion
collisions once the electrons are heated to a temperature $T_e\gtrsim
0.07T_i$. 

The electrons cool by Compton scattering and Bremsstrahlung
emission. Compton scattering dominates for $T_e\gtrsim 10^2$~eV, and
provides a cooling rate of
$\nu_{\rm Comp}=(8\sigma_T/3m_ec)aT_r^4=8T_{r,1.5}^4~{\rm s}^{-1}$.
The electron temperature is determined by
the balance between heating and cooling, 
$\nu_{ie}T_i\approx\nu_{\rm Comp}T_e$, yielding a post-shock electron 
temperature 
\begin{equation}
T_{e+}\approx3\times 10^4 (T_{i,5}\rho_{-10})^{2/5}T_{r,1.5}^{-8/5}~{\rm eV}.
\label{eq:T_e+}
\end{equation}
At this temperature, the ion-electron collision rate is
\begin{equation}
\nu_{ie+}\approx0.4T_{i,5}^{-3/5}\rho_{-10}^{2/5}
T_{r,1.5}^{12/5}~{\rm s}^{-1}.
\label{eq:nu_ie+}
\end{equation}

Based on Eqs. (\ref{eq:nu_ii})--(\ref{eq:nu_ie+}), we conclude that the ion
kinetic energy dissipation is dominated by ion-ion collisions, leading to a
shock front width $\Delta_{s,ii}\approx\nu_{ii}^{-1}v_s
=3\times10^8\rho_{-10}^{-1}$~cm. The electrons are heated by the ions to a
temperature $T_{e+}\approx 30~{\rm keV}$ which is lower by a factor of a
few than the ion temperature, $T_i\approx 100~{\rm keV}$.  Behind the
shock, the electrons cool by Compton scattering, but their temperature is
maintained at $T_e\approx T_{e+}$ over a shell of width
$\approx\nu_{ie}^{-1}v_s\approx10^9\rho_{-10}^{-2/5}$~cm, due to their
coupling to the ions. Over a larger length scale, both the electrons and
ions cool below their post shock temperature.  The electrons maintain a
local thermodynamic equilibrium, since they are heated to
$T_e\gtrsim0.07T_i$ over a length scale $\ll\Delta_{s,ii}$ and subsequently
$\nu_{ee}\gg\nu_{ie}$.

Our results agree with 
the numerical simulations of 
Klein \& Chevalier \cite{Klein789} for a RSG progenitor, who concluded that a
collisional shock is formed at $\tau\sim1$ where 
$T_r\sim10$~eV. 
Klein \& Chevalier derived a post-shock temperature of
$\sim10^4$~eV, 
but pointed out that this result depends on the artificial
viscosity they used, with higher temperatures reached for lower artificial
viscosity values (which yield a sharper shock), and thus concluded that the gas
should be heated up in reality to $T\approx 10^5{\rm eV}$. 
Ensman \& Burrows
\cite{Ensman92} found in their simulations of a BSG progenitor that a
collisional shock forms at $\tau>10$ but reaches lower temperatures, $\sim
1$~keV.  This calculation, however, assumed thermal equilibrium between the
ions and the electrons.  Unfortunately, the resolution of all numerical
simulations so far has not been sufficient to correctly determine the
balance between collisional heating and radiative cooling on the relevant
length scale of $\sim 10^8$ cm.

\paragraph*{Electro-magnetic stability and the collisionless shock.}

The ion velocity distribution within the collisional shock front is highly
anisotropic over a length scale $\Delta_{s,ii}$. This anisotropy leads to
the development of electro-magnetic instabilities. In order to analyze the
shock front stability, we consider a plasma composed of electrons in
thermal equilibrium at a temperature $T_e$, and protons with a
bi-Maxwellian distribution having a temperature $T_\perp=m_p v_s^2$ along
one axis (representing the velocity dispersion associated with the ion
streams perpendicular to the shock front), and a lower temperature in the
orthogonal plane $T_i<T_\perp$ (parallel to the shock front).  Since the
electrons in our case have large thermal speeds, namely
$T_e/T_\perp\sim0.1\gg m_e/m_p$, electrostatic instabilities such as the
two-stream instability, are suppressed. We therefore focus on
electro-magnetic instabilities driven by the ion anisotropy.  Such
instabilities are typically expected to grow at a rate of
$\le(v_s/c)\nu_{pi}$, where $\nu_{pi}$ is the ion plasma frequency (e.g.
\cite{Krall}), with the maximum rate obtained only in the limit of
infinite $T_e$.

Solving the dispersion relation for electro-magnetic waves in the above
plasma, we find purely growing modes (i.e. modes having purely 
imaginary frequencies)
with an electric field along the direction of the high ion temperature
(i.e. perpendicular to the shock front), as long as the wave vector $k$
satisfies the condition $(kc/2\pi\nu_{pi})^2<(T_\perp/T_i)-1$.  In the
limit $m_e/m_p\ll T_e/T_\perp\ll m_p/m_e$, the growth rates are
\begin{equation}
\nu_{EM}\approx\frac{v_s}{c}\nu_{pi} \times
\cases{\min\left(1, k/Y^{1/2}k_*\right),
               &for $T_\perp/T_i\gg Y$;\cr
       \left(\frac{T_\perp}{T_i}-1\right) k/Y^{3/2}k_*, 
               &for $T_\perp/T_i\ll Y$,\cr}
\label{eq:nu_EM}
\end{equation}
where $k_*=2\pi\nu_{pi}/c$ and  $Y=(m_p T_\perp/m_e T_e)^{1/3}$.

For large anisotropy, $T_\perp/T_i\gg Y$, the maximal growth rate is
$(v_s/c)\nu_{pi}$. If the anisotropy is smaller, the maximal growth rate is
$(v_s/c)\nu_{pi}/Y^{3/2}$, which in our case yields
\begin{equation}
\nu_{EM}\approx\left(0.1\frac{m_e}{m_p}\right)^{1/2}\frac{v_s}{c}\nu_{pi}
\approx3\times10^6\rho_{-10}^{1/2}v_{s,9}~{\rm s}^{-1}.
\label{eq:nu_EMN}
\end{equation}
The growth rate of electro-magnetic instabilities is therefore many orders
of magnitude larger than the collision rates $\nu_{ii}$, $\nu_{ie}$.
Correspondingly, the characteristic length scale of the fastest growing
modes, $c\nu_{pi}^{-1}=20\rho_{-10}^{1/2}$~cm, is many orders of magnitude
smaller than the collisional shock width. Under such conditions, we expect
the shock to become collisionless, i.e. to be mediated by plasma
instabilities rather than by collisions. The Ohmic dissipation rate of the
electro-magnetic fields, $\lesssim (m_e/m_p)(\ell/c\nu_{pi}^{-1})^{-2}
\nu_{ee}$, is much lower than their growth rate for all relevant
temperatures and length scales $\ell\gtrsim c\nu_{pi}^{-1}$.

If a collisionless shock indeed forms, it would accelerate charged
particles to high energies, producing a power-low distribution of particle
number, $dN/dp\propto p^{-2}$ where $p$ is the particle momentum
\cite{Eichler87}. The acceleration rate of either relativistic or
non-relativistic protons is given by, $\nu_{\rm acc}\sim\nu_B(E/E_{\rm
th})^{-1}$, where $E$ is the kinetic energy of the accelerated proton,
$E_{\rm th}$ is the energy of thermal protons, and $\nu_B$ is the
(non-relativistic) proton gyro-frequency,
\begin{equation}
\nu_B=\frac{1}{2\pi}\frac{eB}{m_pc}=1.5\times10^4B_1~{\rm s}^{-1},
\label{eq:nu_B}
\end{equation}
where $B_1=(B/10~{\rm G})$ is the magnetic field normalized to 10~G, which
is the typical value inferred at the surface of RSGs \cite{Reid}.  Note,
however, that the electro-magnetic instabilities build-up the magnetic
field.  In analogy with collisionless shocks in supernova remnants
\cite{Helfand}, $\gamma$-ray bursts \cite{Medve}, or the intergalactic
medium \cite{IGM},
the magnetic field may grow up to a sizeable fraction of the equipartition
value, $B_{\rm e.p.}=10^{4.5}\rho_{-10}^{1/2}v_{s,9}$~G, and saturate near
this value due to nonlinear effects \cite{nonlinear}. Thus, the
acceleration rate is likely to significantly exceed the rate implied by
Eq. (\ref{eq:nu_B}).  Since the acceleration rate at thermal energies,
$\nu_B$, is much higher than the collision frequency of thermal protons,
and since the collisional energy loss rate decreases as $E^{-3/2}$ while
the acceleration rate declines only as $E^{-1}$, protons are expected to be
accelerated to energies well above thermal.

Scattering of protons by magnetic field fluctuations is most effective if
the fluctuation wavelength is comparable or larger than the proton Larmor
radius.  The Larmor radius of the thermal protons in an equipartition field
is comparable to the wavelength $c\nu_{pi}^{-1}$ over which instabilities
grow fastest.  As the Larmor radius of highly relativistic protons is much
larger, a conservative lower limit to the maximum energy of the accelerated
protons is given by
\begin{equation}
E_{\rm max}\approx \frac{v_s}{c}eB\Delta=18v_{s,9}B_1\tau^{2/5}~{\rm TeV},
\label{eq:E_max}
\end{equation}
where we have assumed that instabilities do not lead to significant
amplification of the field on large scales. This is probably an
underestimate of the maximum energy, as nonlinear effects are expected to
amplify the field on scales $\gg c\nu_{pi}^{-1}$.  If this occurs, then
protons would be accelerated up to energies well in excess of $10$~TeV.

\paragraph*{Neutrinos and photons.}

Protons accelerated to relativistic energy lose energy by inelastic nuclear
collisions. The average number of nuclear collisions encountered by a
proton is $N_c\approx n\sigma_0c\Delta/v_s=5.2\tau/v_{s,9}$, where
$\sigma_0=50$~mb.  Since a proton loses $\sim20\%$ of its energy in each
collision, the total energy lost to pion production is
\begin{equation}
E_\pi\approx\xi_p\min(1,\tau)\frac{M}{m_p} T_i=
3\xi_p\min(\tau,\tau^2)\times10^{45}{\rm erg},
\label{eq:E_pi}
\end{equation}
where $\xi_p$ is the fraction of the post-shock energy density which is
converted to relativistic protons (note that for a spectrum $dN/dp\propto
p^{-2}$, the energy of super-thermal particles is dominated by relativistic
particles). For the strong shocks around supernova remnants, the inferred
values of $\xi_p$ are of order unity \cite{xip}.

Roughly a third of $E_\pi$ is converted to muon neutrinos, and
a similar fraction to high energy photons, through pion
decay. The neutrino and photon signals should be coincident and
spread over a time scale of $2R_*/c\approx1$ hour.
(The emission from the back
side of the progenitor star is visible for TeV neutrinos but not so for
$\gamma$-rays, and consequently the $\gamma$-ray flash should be shorter by
about a factor of two than the neutrino pulse.)
The predicted flux of $\gamma$-rays and muon neutrinos from a Type
II supernova in our Galaxy is therefore 
\begin{equation}
F_\gamma\approx F_{\nu_\mu}\approx10^{-4}\xi_p\min(\tau,\tau^2)
d_{10\rm kpc}^{-2}~{\rm erg~s^{-1}\,cm}^{-2},
\label{eq:F_gamma}
\end{equation}
where $d_{10\rm kpc}$ is the supernova distance in units of 10 kpc.  The
high density of $\sim10$~eV photons from the radiative shock, will lead to
a high pair-production optical depth for photons with energies $\gtrsim
10$~GeV. Thus, we expect a photon spectrum $dN_\gamma/dE_\gamma\propto
E_\gamma^{-2}$ over the energy range $100{\rm MeV}\lesssim E_\gamma
\lesssim 10{\rm GeV}$. This signal is easily detectable with planned
$\gamma$-ray telescopes (e.g. \cite{GLAST}).

The
neutrino spectrum is expected to follow the accelerated proton spectrum,
with a number per energy interval of $dN/dE\propto E^{-2}$ at relativistic
proton energies. The probability 
that a muon neutrino will produce a high-energy
muon in a terrestrial detector is \cite{Gaisser-rev}
$P_{\nu\mu}\approx1.3\times10^{-6}E_{\nu,\rm TeV}^{\beta}$, with $\beta=2$
for $E_{\nu,\rm TeV}<1$ and $\beta=1$ for $E_{\nu,\rm TeV}>1$. Thus, 
the number flux $J_\mu$ of muon induced neutrinos can be
related to the neutrino energy flux, $F_\nu=\int dE_\nu E_\nu
(dJ_\nu/dE_\nu)$, where $J_\nu$ is the neutrino number flux, through the
relation
\begin{equation}
J_\mu=\int dE_\nu \frac{dJ_\nu}{dE_\nu} P_{\nu\mu}\approx 
\frac{1+\ln(E_{\nu,\rm max}/{\rm 1TeV})}{\ln(E_{\nu,\rm max}/{\rm 1GeV})}
\frac{P_0}{E_0}F_\nu,
\label{eq:J_mu}
\end{equation}
where $P_0/E_0=1.3\times10^{-6}{\rm TeV}^{-1}$. Using Eq. (\ref{eq:E_pi}) we
find that 
\begin{equation}
N_\mu\approx0.15\frac{P_0}{E_0}\frac{E_\pi/3}{4\pi d^2}=
130\xi_p\min(\tau,\tau^2)d_{10\rm kpc}^{-2}{\rm km}^{-2}
\label{eq:N_mu}
\end{equation}
muons are expected to be detected for a single supernova.
The factor 0.15 follows from Eq. (\ref{eq:J_mu}) under the assumption
$E_{\nu,\rm max}\approx1$~TeV. This is probably a conservative estimate, as
proton energies $E_p\gg10$~TeV may be achieved if the magnetic field is
amplified.

\paragraph*{Implications.}

A type II supernova in our Galaxy, which should occur once per 40~years
\cite{Tammann82}, is expected to produce a strong neutrino signal of
$\sim100$ events at $\sim1$~TeV spread over $\sim1$~hr, in planned
square-kilometer neutrino detectors (see, e.g., \cite{Halzen99} for a
review).  
This is the strongest astrophysical signal predicted for the next
generation of TeV neutrino detectors.  
Detection of the neutrino signal
would be feasible also for optically-dark supernovae in the Galactic center
or in molecular clouds, which are heavily obscured by dust.  Combined
analysis of the detected signal along with the thermal (10 MeV, few seconds
long) neutrino burst from the cooling neutron star which should occur about
10 hours earlier ($\sim R_\star/v_s$),
could provide important clues about the formation process of the neutron
star and the dynamics of the resulting supernova shock.

For the neutrino parameters inferred from recent atmospheric experiments
\cite{Atmo-nu}, we expect flavor oscillations to produce equal
fluxes of muon and tau neutrinos upon their arrival to Earth, thus allowing
for a ``$\tau$ appearance'' experiment.  Since the neutrino
signal should coincide with the $\gamma$-ray signal, checking the
simultaneity of photon and neutrino arrival times will allow to test for
deviations from Lorentz invariance and from the weak equivalence
principle. Although the neutrino and photon signals are expected to spread
over $\sim1$~hr, detection of a large number of neutrinos will allow to
test simultaneity to an accuracy much better then 1~hr, thus providing
better limits than those derived from supernova 1987A, where simultaneity
was tested to an accuracy of only several hours (see \cite{Bahcall-book}
for a review).

\paragraph*{Acknowledgments.} 

We thank Z. Barkat, S. Blinnikov, A. Burrows, R. Chevalier, M. Reid and
S. Saar for useful discussions.  This work was supported in part by grants
from the Israel-US BSF (BSF-9800343), MINERVA, NSF (AST-9900877;
AST-0071019), and NASA (NAG5-7039; NAG5-7768). AL thanks 
the Weizmann Institute for its kind hospitality during 
the course of this work.

\end{document}